# LO-phonon assisted polariton laser

M. Maragkou,[a+] A. J. D. Grundy,[a+] T. Ostatnický[a,b] and P. G. Lagoudakis[a,*]

[a]School of Physics and Astronomy, University of Southampton, Southampton, SO17 1BJ, United Kingdom
[b]Faculty of Mathematics and Physics, Charles University in Prague, Ke Karlovu 3, 121 16 Praha 2, Czech Republic

We demonstrate the role of LO-phonon assisted polariton relaxation in reducing the stimulation threshold in strongly coupled microcavities. When the energy of the relaxation bottleneck is one LO-phonon above the ground polariton state, we observe a ten-fold improvement of the polariton relaxation rate in the linear regime, and a two-fold reduction of the threshold to the non-linear polariton lasing regime.

Microcavity polaritons are quasi-bosons that result from the strong-coupling of cavity photons and excitons confined in a single nanostructure.[1] Their bosonic nature gives rise to a plethora of non-linear phenomena[2,3] epitomised by the quantum phase transition to non-equilibrium Bose-Einstein condensation[4,5,6]. The lifetime of polariton condensates is limited to the cavity-photon lifetime leading to the radiation of directional, coherent light, termed polariton laser.[7] Unlike conventional lasers, polariton lasers do not require population inversion and have the potential to operate with reduced thresholds. To date, polariton lasers have only been realised under optical excitation in microcavities with GaAs[8], CdTe[9] and GaN[10] active material. Although electrical pumping in polariton light emitting diodes was achieved both in inorganic and organic microcavities,[11,12,13] polariton lasing remains elusive under electrical carrier injection. As with non-resonant optical excitation schemes, so with electrical injection relaxation bottleneck leads to substantial homogeneous broadening and loss of strong coupling in the linear regime.

In this Letter we demonstrate the role of longitudinal optical (LO)-phonon assisted polariton relaxation in reducing the stimulation threshold in strongly coupled microcavities. The concept of utilising LO-phonon emission to accelerate polariton relaxation was introduced by Imamoglu and co-workers[7] and experimental evidence of the process was shown in the linear regime by Pau et al[14]. Here we tune the energy of the relaxation bottleneck one LO-phonon above the ground polariton state and observe a ten-fold improvement of the polariton relaxation rate in the linear regime, and a two-fold reduction of the threshold for polariton lasing. The introduced configuration exemplifies the potential of LO-phonon emission in appropriately designed polariton diode lasers.

The sample under study is a λ/2 AlAs cavity consisting of two $Al_{0.2}Ga_{0.8}As$/AlAs distributed Bragg reflectors with 16 (top) and 20 (bottom) pairs respectively.[15] Three sets of four quantum wells are located at the centre of the cavity and the first antinodes of the electromagnetic field to increase the Rabi splitting.[16] A wedge in the structure makes accessible a wide range of exciton-photon energy detuning. All experiments are performed at 10 K using a cold-finger cryostat. The sample is excited at the LH energy ($E_{LH}$) with 180 fs p-polarised optical pulses focused to a 40 µm spot.

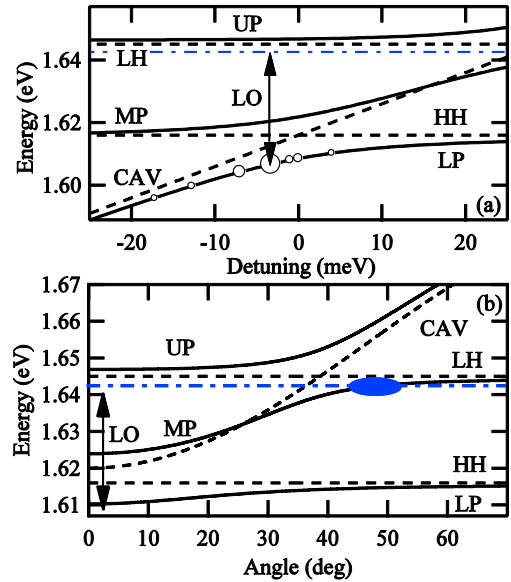

**Fig. 1** Energy dispersion of the cavity and exciton modes (dashed lines) and lower, middle and upper polariton (solid lines) as a function of (a) detuning and (b) angle for a -4meV detuning. The open markers in (a) correspond to experimental data and their size is proportional to the LO phonons scattering strength. The dot-dash blue line (a) and solid blue ellipse (b) indicate the energy of the middle polariton bottleneck.

Following optical excitation free electron-hole pairs rapidly form exciton-polaritons in the middle (MP) and upper (UP) polariton branches, which are subject to further energy relaxation by polariton-phonon and pair polariton scattering. Energy relaxation by longitudinal acoustic (LA) phonon emission predominantly allows for carrier thermalization within the two upper polariton branches, while LO-phonon emission and pair polariton scattering leads to population of the lower polariton (LP) branch below the bottleneck region[17]. The steep dispersion of the middle and lower polariton branches around $k=0$ inhibits LA-phonon emission and results to relaxation bottleneck along the middle branch about 1-3 meV below $E_{LH}$ [Fig. 1(b)]. MP states are mostly light hole (LH)-like at high wavevectors and therefore cannot effectively relax to the heavy hole (HH)-like LP by LO-phonon emission, the relaxation rate is limited by the weak electron-hole exchange interaction. At the MP bottleneck region strong exciton-photon coupling mixes both LH and HH excitons rendering LO-phonon emission the strongest energy relaxation mechanism. The presence of the MP bottleneck provides a polariton reservoir from which LP states can be effectively populated by LO-phonon emission.

We excite resonantly to the light-hole (LH) exciton energy, record photoluminescence from the ground state of the lower-polariton (LP) branch and tune the energy difference between the three polariton branches by scanning the excitation spot across the wedged cavity LP [Fig. 1(a)]. At each detuning the excitation intensity is scaled for the reflectivity of the stop-band to accurately control carrier excitation density. The detuning dependence of the ground polariton state is shown in Fig. 1(a). The open markers correspond to the LP emission energy in the linear regime. Their size is scaled with the LO phonon transition rate, discussed later. The MP bottleneck energy is indicated by the blue dot-dash line. The calculated angular dispersion for the optimum detuning (-4meV) is shown in Fig. 1(b), where the MP relaxation bottleneck region is one LO-phonon energy, $E_{LO}$, above the LP ground state.

fitted with a straight line (a), whose slope can give information on the carrier relaxation efficiency. Strong coupling is confirmed by the below (d) and above (e) threshold snapshots of the dispersion relation.

Time-integrated photoluminescence from the LP ground state is recorded with ±3° collection angle and is spectrally resolved using a 1200 grooves/mm grating in a 55 cm spectrometer coupled to a cooled CCD. A non-linear increase of the photoluminescence intensity by $10^2$ at threshold is observed with increasing excitation density [Fig. 2(a)]. Polariton lasing occurs at threshold as the polaritons collapse to a single state. Self interaction results in a blueshift of the photoluminescence by ~3meV [Fig. 2(b)] and an increase in coherence causes a collapse of the linewidth [Fig.2(c)].[18] Under pulsed excitation the blueshift of the LP evolves with time. This information is lost in time-integrated measurements and the dominance of polariton lasing results in an artificially large jump in mode energy upon reaching threshold [Fig. 2(b)]. These observations are in accordance with previous reports on polariton lasing[19]. Figures 2(d,e) show snapshots of the LP dispersion in the linear regime and at threshold, respectively.

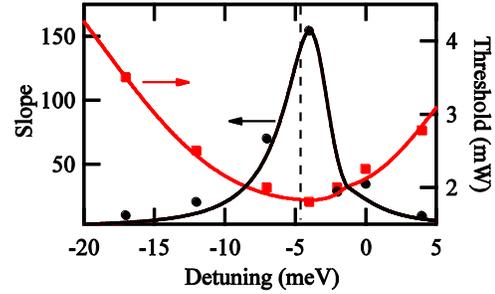

**Fig. 3** Experimental data (markers) and theoretical fits (lines) for the below threshold slope $S(D)$(black) and the stimulation threshold $T(D)$(red) for a range of detunings.

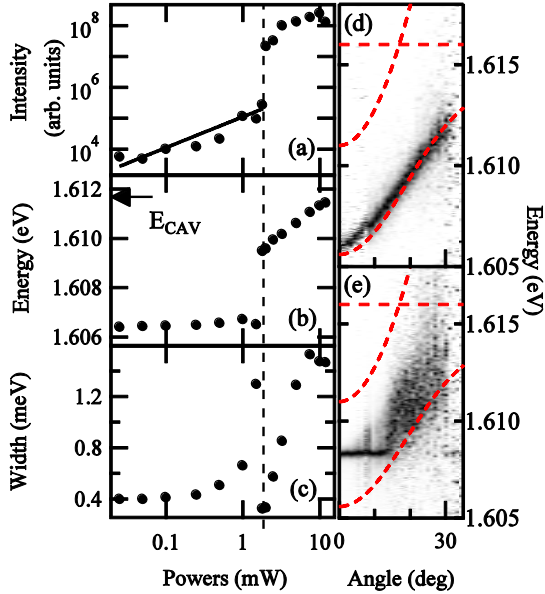

**Fig. 2** Power dependence of (a) the emission intensity, (b) the energy blueshift and (c) the linewidth of the lower polariton ground state as a function of the excitation power for -4meV detuning. The below threshold power dependence data can be

As discussed above, by changing the cavity-exciton detuning (D), the energy gap between the MP bottleneck and the LP ground state can be tuned through the $E_{LO}$ resonance. This leads to a detuning dependence of the LO phonon transition rate $W(D)$, which is maximum at resonance. The transition rate $W(D)$ can be found from the slope $S$ of the power dependence in the linear regime (left of the dashed line in Fig. 2(a)). A greater slope indicates an increased transition rate: $S(D) \propto P_{LP}^2(D)W(D)$, where $P_{LP}(D)$ is the photon Hopfield coefficient of the $k=0$ LP ground state. Fig. 3 shows experimental (black dots) data for the detuning dependence of $S(D)$. It can be seen that the slope is almost two orders of magnitude greater when the LO phonon resonance condition is met, at a detuning of -4meV, compared to the off-resonance condition of

+4meV or -13meV. The black line in Fig. 3 is a theoretical fit calculated using Fermi's golden rule for the relaxation rate $W(D)$, assuming that the MP bottleneck occurs 2meV below the LH and that the polariton line broadening is 1.3meV (taken from Fig. 2(c)).

The stimulation threshold $T$ is also dependent on $W(D)$. An increased transition rate means that the ground state is populated more efficiently, allowing the macroscopic population required for stimulation and lasing to occur at reduced pump powers. Experimental data for the detuning dependence of the threshold to polariton lasing is shown in Fig. 3 (red dots). The stimulation threshold is assumed to have the form $T(D) = \alpha + \beta / W(D)$, where $\alpha$ and $\beta$ are constants chosen to best fit the experimental data. It shows that the threshold is a minimum at the detuning where the LO phonon transition is on resonance (-4meV), 50% lower than the off resonance case. The red line is a fit to the data based on the simple model and assumptions described earlier.

Further studies are performed with the pump laser tuned to be non-resonant. In this case we find similar trends to those under resonant excitation. The optimum detuning remains at -4meV, where the lowest threshold and highest slope are achieved. Interestingly the threshold (scaled with the absorption coefficient) is similar for both excitation configurations, suggesting that LO phonons mediated relaxation mechanisms are present in both pumping schemes. However, the slope of the power dependence in the linear regime is about 3 times lower compared to resonant excitation. We suggest that under non resonant excitation, a smaller proportion of the injected carriers collect at the MP bottleneck and undergo LO phonon relaxation. In this case, it is likely that more carriers undergo LA phonon relaxation to the LP, which would not be desirable for electrical injection.

Using an LO phonon transition to reduce the polariton stimulation threshold may prove significant in the development of electrically pumped polariton lasers. We propose a new design for an electrically injected polariton laser where carriers are injected at or above the $E_{LO}$ resonance. Resonant injection allows LO-phonon mediated relaxation to dominate over LA-phonon, bypassing the bottleneck effects that have prevented the operation of electrically pumped polariton lasers to date. In order to realize this we propose that samples are designed with the appropriate detuning to make the MP bottleneck $E_{LO}$ above the ground polariton state. Carriers should be injected at, or above the MP branch energy, allowing the MP bottleneck to provide a reservoir from which LO-phonon scattering to the ground polariton state can occur. We have demonstrated this in GaAs based systems and propose this to be the material of choice for polariton laser devices, as opposed to the InGaAs quantum well microcavities.

In conclusion, we have achieved a 50% reduction in the polariton stimulation threshold in a 2D GaAs microcavity by using LO phonons to efficiently relax polaritons from the MP bottleneck to the LP ground state. We have studied the LO phonon transition rate using power dependences in the linear regime and found that relaxation is more efficient when carriers are injected resonantly with the LH compared to non-resonant injection. We suggest that this mechanism could be used to overcome the relaxation bottleneck which hinders electrically injected polariton lasers. Finally we propose a design that uses LO-phonon relaxation to bypass the relaxation bottleneck in an electrically injected polariton laser.

P.G.L would like to acknowledge EPSRC grant EP/F026455/1 and FP7 grants ITN-Clermont 4 and ITN-Spinoptronics for funding. This work is also part of the research plan MSM0021620834 that is financed by the Ministry of Education of the Czech Republic. The authors thank Jacqueline Bloch and Lemaitre Aristide for provision of the sample and Elena Kammann for useful discussions.